\documentclass[twocolumn,aps,showpacs,amsmath,amssymb,floatfix,superscriptaddress]{revtex4}
\usepackage{graphicx}
\usepackage{dcolumn}
\usepackage{bm}
\usepackage{array}
\usepackage{hyperref} 
\usepackage{latexsym}
\usepackage{float}
\begin{document}
\title{Aging and Immortality in a Cell Proliferation Model}
\author{T. Antal} 
\affiliation{Center for Polymer Studies and Department of Physics, Boston
  University, Boston, MA, USA 02215}
\affiliation{Program for Evolutionary Dynamics, Harvard
  University, Cambridge, MA, USA 02138}

\author{K.B. Blagoev}
\author{S.A. Trugman}
\affiliation{Theory Division, Los Alamos National Laboratory, Los
  Alamos, New Mexico, USA 87545}
\author{S.~Redner}
\affiliation{Center for Polymer Studies and Department of Physics, Boston
  University, Boston, MA, USA 02215}

\begin{abstract}
  
  We investigate a model of cell division in which the length of telomeres
  within a cell regulates its proliferative potential.  At each division,
  telomeres undergo a systematic length decrease as well as a superimposed
  fluctuation due to exchange of telomere DNA between the two daughter cells.
  A cell becomes senescent when one or more of its telomeres become shorter
  than a critical length.  We map this telomere dynamics onto a biased
  branching diffusion process with an absorbing boundary condition whenever
  any telomere reaches the critical length.  Using first-passage ideas, we
  find a phase transition between finite lifetime and immortality (infinite
  proliferation) of the cell population as a function of the influence of
  telomere shortening, fluctuations, and cell division.

\end{abstract}

\maketitle

\section{Introduction}

Aging is a complex and incompletely understood process characterized by
deteriorating cellular, organ, and system function.  Replicative senescence,
the phenomenon whereby normal somatic cells show a finite proliferative
capacity, is thought to be a major contributor to this decline
\cite{general}.  Telomeres are repetitive DNA sequences ((TTAGGG)$_n$ in
human cells) at both ends of each linear chromosome and their role is to
protect the coding part of the DNA.  Normal human somatic cells become
senescent after a finite number of doublings.

As a general rule, cells for which telomerase activity is absent lose of the
order of 100 base pairs of telomeric DNA from chromosome ends in every cell
division.  This basal loss has been attributed to the end replication problem
\cite{Watson} in which the DNA-polymerase cannot replicate all the way to the
end of the chromosome during DNA lagging strand synthesis.  Additional
post-replicative processing of the telomeric DNA is necessary to protect the
end of the chromosome from being recognized as a double strand break in need
of repair.  This processing also contributes to the end replication problem.

What remains unexplained, however, is why senescent cells occur in cell
cultures long before the expected number of cell divisions estimated from the
gradual basal loss.  It has been shown recently that in addition to the basal
loss of $\sim 100$ base pairs per division, a complex set of events that
leads to telomere exchange between sister chromatids can occur.  This
telomere sister chromatid exchange (T-SCE), together with basal telomere loss
and a number of observed or suggested telomere recombination events,
collectively define telomere dynamics, and this dynamics leads to a wide
distribution of telomere lengths in cell cultures \cite{TCE}.  Telomere
exchange can also occur between the telomeres of different chromosomes.
Currently available data cannot distinguish between this process and T-SCE.
It is also believed that sister chromosome exchange is induced by DNA damage.
Because the sister chromatids are in closer proximity compared to the
distance between different chromatids, it can be hypothesize that the
probability for telomere interchromatid exchange (T-ICE) is smaller than the
probability for sister chromatid exchange between sister chromatids.

Recently, one of the present authors proposed a theory \cite{BlagoevGoodwin},
based on telomere dynamics including T-SCE and T-ICE, that is capable of
explaining Werner's syndrome, an inherited disease characterized by premature
aging and death, and a subset of cancers that seem to use a recombination
mechanism to maintain telomere length.  This latter mechanism, known as
alternative lengthening of telomeres (ALT), as well as additional telomerase
activity in some cases, are thought to contribute to the large proliferative
potential of these cells \cite{ALT}.  Here proliferative potential denotes
the time over which cells can continue to divide.  In
Ref.~\cite{BlagoevGoodwin}, we showed that both Werner's syndrome and ALT can
be described within the general framework of telomere dynamics in which there
is an elevated rate of exchange of telomere DNA between two daughter cells,
as observed in many experiments \cite{expt}.  One of the main numerical
results from Ref.~\cite{BlagoevGoodwin} is a transition from finite to
infinite proliferative potential in a cell culture as the parameters
controlling the telomere recombination rates are varied.

\begin{figure}[ht]
\includegraphics*[width=0.275\textwidth]{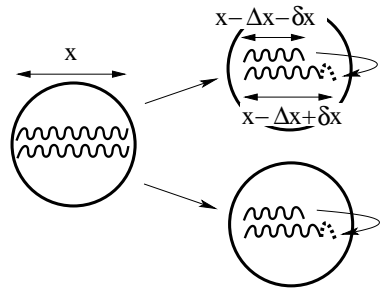}
\caption{Schematic illustration of telomere evolution.  The telomeres in the
  initial cell contain $x=15$ units.  Upon division, each telomere ostensibly
  shortens by $\Delta x=5$ units, but additional exchange of $\delta x=\pm 2$
  units between daughter telomeres (dashed) leads to final lengths of 8 and
  12.}
\label{cartoon}
\end{figure}

In this work, we analytically investigate an idealized model of cell
proliferation in which the number of cell divisions before senescence occurs
is controlled by the dynamics of telomeres during cell division.  Each cell
contains a certain number of telomeres.  When a cell divides, the telomeres
in each daughter cell ostensibly shorten by a fixed amount $\Delta x$.  In
addition to this systematic telomere shortening, the effect of T-SCE
processes during cell division leads to a superimposed stochastic component
to the telomere length dynamics by an amount $\delta x$ (Fig.~\ref{cartoon}).
Thus in each cell division event, the length of an individual telomere
evolves by the combined effects of these systematic and stochastic mechanisms.

When the length $x$ of any of the telomeres within a cell reaches zero, the
cell stops dividing and becomes senescent.  On the other hand, the stochastic
component of the telomere dynamics provides the possibility for a telomere to
occasionally grow when a cell divides.  This sub-population of cells with
long telomeres and thus higher proliferative potential can become even more
so at the next cell division, a mechanism that allows a long-lived
subpopulation to thrive.


We are interested in basic statistical properties of the cell proliferative
potential.  Some fundamental questions that we will study include:
\begin{enumerate}
\item Can a cell population divide indefinitely?  
\item How long does it take for a cell population to become senescent?
\item How many dividing cells exist after a given number of divisions?  
\end{enumerate}
Within an idealized model of telomere kinetics described by Eq.~\eqref{def}
for a single telomere per cell, it is possible to answer these questions
analytically by mapping the telomere dynamics to a first-passage process.
Using this approach, we find a phase transition between a finite-lifetime
cell population and immortality as a function of three basic control
parameters---the magnitude of the systematic part of the telomere evolution,
determined by $\Delta x$, the effective diffusion coefficient associated with
the stochastic part of the telomere evolution, determined by $\langle \delta
x(t)^2\rangle$, and the cell division rate.

\section{Telomere Replication Model}

In our telomere replication model, we assume that the initial length of each
telomere in a cell is $x_0$.  In any cell division event, the length of a
telomere changes by two distinct processes:
\renewcommand{\theenumi}{\roman{enumi}}
\begin{enumerate}
\item a systematic shortening of each telomere by $\Delta x$;
\item an additional stochastic component of the length change of magnitude
  $\delta x$.
\end{enumerate}
Thus the length of a telomere changes according to
\begin{eqnarray}
\label{def}
x\to x-\Delta x+\delta x.
\end{eqnarray}
Here the stochastic variable $\delta x$ accounts for the T-SCE processes that
we assume to have mean value equal to zero, $\langle\delta x\rangle=0$, and
no correlations at different times, $\langle\delta x(t)\delta
x(t')\rangle\propto \delta(t-t')$.  The justification for the absence of correlation
is that T-SCE events have been linked to DNA damage \cite{TCE-justify}, which
occurs randomly in the cell.

Because of the systematic and stochastic contributions to the change in
telomere length in each division event, the length of a telomere undergoes a
biased random walk, with a bias toward shrinking.  It is instructive to
estimate the relative importance of the systematic and stochastic components
of this length evolution.  For this purpose, we define the time unit as the
physical time between cell divisions $\delta t$.  Thus in the absence of
stochasticity a telomere shrinks to zero length in $x_0/\Delta x$ cell
division events.

It is now helpful to recall some basic numbers about human telomeres to
connect our mechanistic telomere model and real cell division: \bigskip

\begin{tabular}{|l|l|l|}
  \hline
  parameter &  definition    &   numerical value \\
  \hline
  $\delta t$& time between & 20 minutes -- \\
  & cell divisions& several hours\\
  \hline
  $x_0$   &initial telomere &  $\approx 10^4$ base pairs  \\
  & length & \\
  \hline
  $\Delta x$   &systematic length   & $\approx 10^2$ base pairs  \\
  & decrease per division & \\
  \hline
  $\delta x$   &stochastic length  & $\approx 10^2$ base pairs  \\
  & decrease per division &\\
  \hline
\end{tabular}
\medskip

\noindent The quantity $\delta x$ is known only very roughly.  It is
possible, with low probability, that even whole telomeres can be lost in TCE
processes \cite{low-prob}.

With the above numbers, a telomere shrinks to zero in $x_0/\Delta x\approx
10^2\equiv N$ cell division events with purely deterministic shrinking.  Now
consider the role of stochasticity: in $N$ cell divisions, the
root-mean-square length change due to stochastic events is $ \ell_{\rm rms}
=\sqrt{N(\delta x)^2} \approx 10^3$.  Thus in the time for a telomere to
systematically shrink from $10^4$ to zero, stochasticity gives a length
uncertainty of $10^3$---a 10\% correction to the bias.

Finally, we need to include the role of cell division on this biased random
walk description to arrive at a theory for telomere dynamics.  That is, we
need to allow a random walk to replicate as it undergoes biased hopping.
While there are many ways to parameterize the effects of bias, stochasticity,
and replication in the continuum limit, all such models lead to the following
convection-diffusion equation with multiplicative growth:
\begin{equation}
\label{diff1}
\frac{\partial n(x,t)}{\partial t} = kn(x,t) + v\frac{\partial
  n(x,t)}{\partial x} + D \frac{\partial^2 n(x,t)}{\partial x^2}\,,
\end{equation}
in which $v$ represents the bias for telomere shrinking, $D$ accounts for the
stochastic part of the telomere length evolution, and $k$ accounts for cell
division.  While there is only an indirect connection between the model
parameters $v,D,k$, and the parameters $\Delta x, \delta x$ that account for
what happens to a telomere in a single cell division, this continuum
description has the advantage of capturing the physical essence of telomere
dynamics while being analytically tractable.

The basic question that we seek to understand is how long it takes for a cell
to become senescent, an event that occurs when the length of one of its
telomeres reaches zero.  This condition translates to an absorbing boundary
condition at $x=0$ for the biased diffusion process that described the
telomere length distribution.  We now exploit some classic results about the
first-passage probability of biased diffusion (see Appendix and
\cite{fpp}) to determine the evolution of the telomere length
distribution.  Let the initial number of cells be $N_0$.  The solution of
\eqref{diff1} is simply
\begin{equation}
\label{n1}
 n(x,t) = N_0 e^{kt} c(x,t)
\end{equation}
where $c(x,t)$, given in \eqref{cxt}, is the solution to the
convection-diffusion equation with an absorbing boundary condition at $x=0$
and the initial condition that all cells have initial telomere length $x_0$.

We may also treat the situation in which each cell contains $M$ {\it
  independent\/} telomeres as a zeroth-order description for cells that
contain many telomeres whose dynamics is coupled by the T-SCE exchange
process (schematically illustrated in Fig.~\ref{cartoon}).  This assumption
of independence of different telomeres allows us to apply the single telomere
per cell first-passage description with only minor modifications.  For cells
containing $M$ independent telomeres of lengths ${\mathbf x} = (x_1,\dots,
x_M)$, the density of cells $n({\mathbf x}, t)$ satisfies the $M$-dimensional
convection-diffusion-growth equation
\begin{equation}
\label{diffM}
\frac{\partial n({\mathbf x},t)}{\partial t} = kn({\mathbf x},t) + 
v\nabla n({\mathbf x},t) + D \nabla^2 n({\mathbf x},t) ~,
\end{equation}
with absorbing boundaries when any telomere length $x_i$ reaches 0.  The
solution simply factorizes as a product of one-dimensional solutions
\begin{equation}
\label{nM}
 n({\mathbf x},t) = N_0 e^{kt} \prod_{i=1}^M c(x_i,t) ~.
\end{equation}
If each telomere has initial length $x_0$, all the $c(x_i,t)$ are the same
and are given by \eqref{cxt}.  We can also straightforwardly study the case
where each telomere has a different initial length by merely using the unique
initial length of each telomere in Eq.~\eqref{nM}.  It is worth mentioning
that in addition to its application to cell division statistics,
Eq.~\eqref{diffM} is also related to the Fleming-Viot (FV) process \cite{FV}
in which a population of diffusing particles can get absorbed at boundary
point and then be re-injected into the system at a rate that is proportional
to local particle density.

From our description of telomere dynamics as a biased branching-diffusion
process, we now determine basic features about the time dependence of the
cell population and the statistics of telomere lengths.

\subsection{Number of Dividing and Senescent Cells}

Cells in which each telomere has positive length can divide.  The number of
such active cells is given by the integral of the number density of cells
over the positive $2^M$-tant ($x_1, x_2,\ldots, x_N>0$) of length space:
\begin{equation}
\label{N}
N_{\rm active}(t) = \int _{{\mathbf x}>0} n({\bf x},t)\, d{\mathbf x} = N_0
e^{kt} S^M(t) ~,
\end{equation}
where $S(t)$ is the survival probability of a biased random walk (given by
\eqref{S}).  From \eqref{Slarget}, the long-time behavior of $N_{\rm
  active}(t)$ is given by
\begin{eqnarray}
\label{Nt1}
N_{\rm active}(t) &\sim& \Big(\sqrt\frac{D}{\pi}\frac{2x_0}{v^2}\Big)^{\!\!M}\!\!
 e^{Mvx_0/2D}~ t^{-3M/2}~ e^{(k-Mv^2/4D)t}\,,\nonumber \\
&\propto&   t^{-3M/2}~ e^{kt(\epsilon_M-1)/\epsilon_M}\,,
\end{eqnarray}
with $\epsilon_M \equiv 4Dk/Mv^2$.  Thus the fundamental parameters of the
system are the P\'eclet number $P\!e \equiv vx_0/2D$ \cite{Pe}, a
dimensionless measure of the relative importance of the bias and the
stochasticity and, more importantly, the dimensionless growth rate,
$\epsilon_M$.  Note that both the diffusion coefficient $D$ and the bias
velocity $v$ are proportional to $k$, since the telomere length changes occur
only when a cell divides.  As a result, the growth rate $\epsilon_M$ is
actually independent of $k$.  For $\epsilon_M<1$, cell division is
insufficient to overcome the effect of inexorable death due to the systematic
component of the telomere shortening and the population of dividing cells
decays exponentially in time.

\begin{figure}[ht]
\includegraphics*[width=0.375\textwidth]{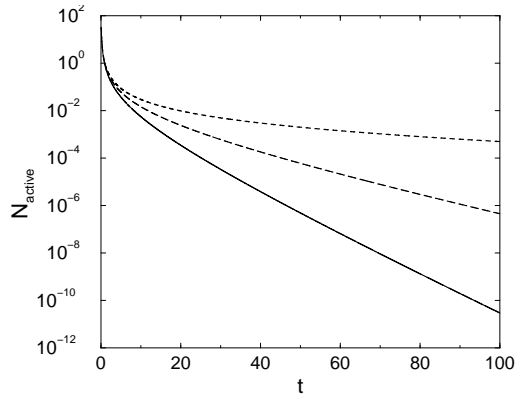}
\caption{Plot of the number of active cells versus time from the second line
  of Eq.~\eqref{Nt1} for the case of $M=1$ telomere per cell, with
  $\epsilon_1=0.8$ (solid) 0.9 (dashed), and 0.99 (dotted).}
\label{fig-Nactive}
\end{figure}

To get a feeling for the interplay between telomere shortening and cell
division, let us employ the numerical parameters given at the beginning of
this section in Eq.~(\ref{Nt1}) to obtain the total number of dividing cells.
Since cells double at each time step, $k=\ln 2$ when we express $t$ in units
of cell division times.  Furthermore, we define the coefficients $\alpha$ and
$\beta$ by $v=\alpha\times 10^2$ and $D=\beta\times 10^4$.  Then the time
dependent exponential factor in (\ref{Nt1}) for the case $M=1$ becomes
\begin{eqnarray*}
e^{(k-v^2/4D)t}\sim e^{(\ln 2-\alpha^2/4\beta)t}\,.
\end{eqnarray*}
The exponent can be either positive or negative exponent depending on
$\alpha$ and $\beta$, which, in turn, depend on details of the telomere
evolution a single cell division event.  Thus, using the numbers appropriate
for humans, senescence or immortality is controlled by details of telomere
evolution as encoded by the coefficients $\alpha$ and $\beta$.

We also obtain the total number of cells that become senescent at any given time during
the evolution as the diffusive flux to $x_i=0$ for any $i$.  For the
symmetric initial condition, the number of cells reaching any boundary
$x_i=0$, $i=1,2,\ldots,M)$, are the same.  Hence we may consider a single
boundary, say $i=1$.  The number of dying cells at this boundary can be
written as the integral over the $x_1=0$ surface
\begin{equation}
J_1(t) = \int_{{{\mathbf x}>0}\atop x_1=0} D\, \frac{\partial n({\mathbf x}, t)}{\partial x_1} 
\,  d{\mathbf x}_\perp\,,
\end{equation}
where the integral is over all the $M-1$ coordinates perpendicular to
$x_1$.  Note again the absence of a convective term in this expression
because there is no convective flux when the concentration is zero.  Using
the product form \eqref{nM} of $n({\mathbf x}, t)$ for the total number of
dying cells we obtain
\begin{equation}
\label{J}
 J(t) = MJ_1(t) = N_0 M e^{kt} F(t) S^{M-1}(t)
\end{equation}
where $S(t)$ and $F(t)$ are given by \eqref{S} and \eqref{F} respectively.
The total number of senescent cells that are produced during the course of the
evolution is
\begin{eqnarray}
N_{\rm sen}&=&\int_0^\infty J(t)\, dt =
- N_0 \int_0^\infty dt~ e^{kt}~ \frac{\partial  S^M(t)}{\partial t}\nonumber \\ 
&=& N_0 \left[ 1- k \int_0^\infty dt~ e^{kt}~ S^M(t)  \right] ~,
\end{eqnarray}
where we again use the fact that $F(t)=-d S(t)/dt$ to perform the integration
by parts.

\begin{figure}[ht]
\includegraphics*[width=0.375\textwidth]{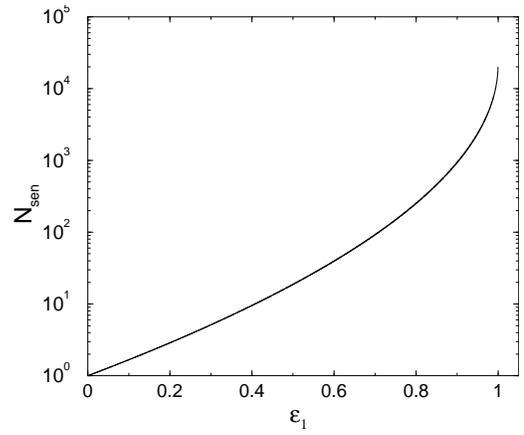}
\caption{Plot of the total number of senescent cells that are produced during
  the evolution of the cell population as a function of the dimensionless
  growth rate $\epsilon_1 =4Dk/v^2$.}
\label{fig-Nsen}
\end{figure}

For the special case of one telomere per cell ($M=1$) we have
\begin{equation}
J(t)= \frac{N_0 x_0}{\sqrt{4\pi Dt^3}}\,\, e^{kt}\,\,e^{-(x_0-vt)^2/4Dt}.
\end{equation}
In this case the total number of senescent cells is
\begin{equation}
N_{\rm sen}=
\frac{N_0 x_0}{\sqrt{4\pi D}}\, \int_0^\infty \frac{1}{t^{3/2}}\,\,e^{kt}\,\,
e^{-(x_0-vt)^2/4Dt} dt.
\end{equation}
We now make the substitution $z=t^{-1/2}$ to recast the above integral into
the form
\begin{equation}
N_{\rm sen}= \frac{N_0 x_0}{\sqrt{\pi D}}\, e^{vx_0/2D}
\int_0^\infty dz\,\, e^{-x_0^2z^2/4D-(\frac{v^2}{4D}-k)/z^2}.
\end{equation}
Using the $\int_0^\infty e^{-az^2-b/z^2}\, dz = \sqrt{\frac{\pi}{4a}}\,
e^{-\sqrt{4ab}}$ from 3.325 of Ref.~\cite{GR}, we thus obtain
\begin{eqnarray}
\label{Nsen}
N_{\rm sen}&=&N_0 e^{vx_0/2D}\,\exp\left[ -\sqrt{\frac{v^2x_0^2}{4D}\Big(1-\frac{4Dk}{v^2}\Big)}\right]
\nonumber \\
&=&  N_0 e^{P\!e}\, \exp\left[-\sqrt{P\!e^2(1-\epsilon_1)}\right],
\end{eqnarray}
which is plotted in Fig.~\ref{fig-Nsen} for the case $N_0=1$ and $P\!e=10$.

This result for $N_{\rm sen}$ holds only for $\epsilon_1<1$; this is the
regime where the cell population eventually becomes senescent so that the
total number of senescent cells produced during the evolution is finite.
When $\epsilon_1\ll 1$, the leading behaviors of the two exponential factors
in Eq.~(\ref{Nsen}) cancel and $N_{\rm sen}\to (1+\epsilon_1/2P\!e)$.  Thus
as $\epsilon_1\to 0$ (no division), the initial cell immediately becomes
senescent so that the total number of senescent cells that are produced
equals one.  This unrealistic result arises because in the continuum
description telomeres can shrink to zero length before any cell division can
occur.  On the other hand, for $\epsilon_1>1$, the population is immortal and
an infinite number of senescent cells are produced during the evolution.

\subsection{Telomere Length Distribution}

A curious aspect of the population of dividing cells is the dependence of the
cell density on telomere length as $t\to\infty$.  For any number of telomeres
per cell $M$, the telomere length distribution is independent of this number,
since the distribution is simply proportional to $c(x,t)$ in Eq.~(\ref{cxt}).
In the $t\to\infty$ limit, we then obtain
\begin{eqnarray}
\label{c-asymp}
c(x,t\to\infty)&\propto& \frac{xx_0}{\sqrt{4\pi (Dt)^{3}}}\,  e^{-(v^2/4D)t-v(x-x_0)/2D}
\nonumber \\
&\propto& x e^{-vx/2D} ~.
\end{eqnarray}
Thus apart from an overall time-dependent factor, the density of dividing cells
in which the constituent telomere has length $x$ is linear in $x$ for small
lengths and has an exponential cutoff for large lengths.  From the
distribution given in Eq.~\eqref{c-asymp}, the mean telomere length goes to a
constant for large times
\begin{equation}
 \langle x\rangle = \frac{\int _0^\infty x\, c(x,t)\, dx}{\int _0^\infty c(x,t)\, dx}
 \to \frac{4D}{v} ~,
\end{equation}
independent of whether the total cell population is growing or decaying.  The
variance of the telomere length also approaches a constant
\begin{equation}
 \langle x^2\rangle - \langle x\rangle^2 
 \to \frac{8D^2}{v^2} ~.
\end{equation}

A more pedantic way to arrive at this same result is to compute the exact
time-dependent mean telomere length $\langle x(t)\rangle$ using
Eq.~\eqref{cxt} for $c(x,t)$ and then taking the $t\to\infty$ limit.  In
summary, the mean telomere length is time independent, as predicted by the
asymptotic form of the telomere length distribution in Eq.~(\ref{c-asymp}).
We emphasize that this result pertains to the fraction of cells that are
active.  If all cells---senescent and active---are included in the average,
then $\langle x\rangle$ would asymptotically decay with time.

\subsection{Mean Proliferative Potential and Immortality}

As discussed above, for $\epsilon_M<1$, all cells eventually become
senescent.  However, for $\epsilon_M>1$, a subpopulation of infinitely
dividing cells arise so that the average cell population becomes immortal.
We make this statement more precise by computing the average lifetime of the
population for $\epsilon_M<1$.  This lifetime is defined as the average age
of each cell when it becomes senescent and is thus given by
\begin{eqnarray}
\label{tav-def}
\langle t\rangle =\frac{\displaystyle \int_0^\infty t\, J(t)\, dt} {\displaystyle\int_0^\infty  J(t)\, dt}
=  \frac{\displaystyle \int_0^\infty t\,e^{kt} F(t) S^{M-1}(t)  \, dt} 
{\displaystyle\int_0^\infty e^{kt} F(t) S^{M-1}(t)\, dt} ~,
\end{eqnarray}
where we use \eqref{J} for $J(t)$, the number of cells that become senescent
at time $t$.  More simply, the mean proliferative potential (lifetime) can
also be obtained from the total number of senescent cells via $\langle
t\rangle = \frac{\partial}{\partial k} \ln N_{\rm sen}$.

Using the general definition of Eq.~\eqref{tav-def}, it is straightforward to
compute higher moments of the lifetime for the case $M=1$.  Here the
integrals in \eqref{tav-def} can be written in terms of the modified Bessel
functions of the second kind (see \#12 in 3.471 of Ref.~\cite{GR}) to give
\begin{equation}
 \langle t^\gamma \rangle = \left( \frac{x_0^2}{v^2(1-\epsilon_1)} \right)^{\gamma/2}
 \frac{K_{\gamma-1/2} (\sqrt{1-\epsilon_1}Pe/2)}{K_{-1/2} (\sqrt{1-\epsilon_1}Pe/2)} ~,
\end{equation}
where $\gamma$ does not necessarily have to be an integer.  Using this
general formula the variance of the lifetime is
\begin{equation}
  \langle t^2\rangle - \langle t\rangle^2 = \frac{2Dx_0}{v^3(1-\epsilon_1)^{3/2}}~.
\end{equation}
In fact, the $n^{\rm th}$ cumulant can be obtained simply from the general
formula \cite{K}
\begin{eqnarray*}
C_n = \frac{\partial^n \ln N_{\rm sen}}{\partial k^n}.
\end{eqnarray*}
Therefore for the higher moments of the lifetime, the diffusion coefficient
does play an essential role. 

For $M=1$, we immediately obtain from Eq.~\eqref{Nsen}
\begin{equation}
\label{tav}
\langle t\rangle = \frac{x_0}{v}\, \frac{1}{\sqrt{1-\epsilon_1}}.
\end{equation}
Thus the mean cell proliferative potential diverges to infinity as
$\epsilon_1\to 1$ from below.  This result for the number of cell divisions
before senescence is one of our primary results.  Notice that in the case of
no cell division ($k=0$, $\epsilon_1=0$) the average lifetime $\langle
t\rangle = x_0/v$.  That is, $\langle t\rangle$ coincides with the time for a
biased diffusing particle to be convected to the origin.  It is surprising at
first sight that diffusion plays no role in determining the average number of
cell divisions.  Exactly the same type of result arises for the discrete random
walk with a bias $v$ \cite {kt}.

\subsection{Senescence Plateau}

A useful characterization of the number of cell divisions distribution
function of a population is the {\it senescence rate} $m(t)$.  The senescence
(or mortality) rate is the ratio of the number of cells that become senescent
at time $t$ to the total number of cells that are still dividing at this
time.  Equivalently, the senescence rate is the probability that a randomly
chosen dividing cell becomes senescent at the next moment.  The senescence
rate is thus given by
\begin{equation}
\label{mdef}
 m(t) = \frac{J(t)}{N(t)} ~,
\end{equation}
where the number of dying cells $J(t)$ and dividing cells $N(t)$ are given by
Eqs.~\eqref{J} and \eqref{N} respectively.  Substituting these expressions
into \eqref{mdef}, we find that the senescence rate is independent of the
number of telomeres per cell $M$, {\it i.e.}
\begin{equation}
 m(t) = \frac{F(t)}{S(t)} = -\frac{\partial \ln S(t)}{\partial t} ~.
\end{equation}

Using the asymptotic form of $S(t)$ given in \eqref{Slarget}, the senescence
rate approaches a time-independent value in the long-time limit and is given
by
\begin{equation}
 m(t) \simeq \frac{v^2}{4D} + \frac{3}{2t} + {\cal O}\left(\frac{1}{t^2}\right) ~.
\end{equation}
Amazingly, as the cell population ages, the senescence rate of the cells that
remain dividing ultimately tends to a constant value for large times.  This
phenomenon is known as the {\it mortality plateau}.  Namely, the probability
that the somatic cells of an organism become senescent becomes {\it
  independent\/} of its age in the long-time limit.  This surprising fact was
observed experimentally in human populations \cite{humans} and for fruit
flies \cite{flies}.  It was also observed numerically in a model of aging
that is similar to ours \cite{WF}.  The
existence of such a senescence plateau is actually typical of a wide range of
Markov processes \cite{evans}.

\section{Summary}

We studied an idealized model for the dynamics of telomere lengths during
cell division that is based on a systematic basal loss and a stochastic
component to the evolution that arises from telomere sister chromatid
exchange.  This model captures essential features of telomere dynamics in
cell cultures.  Because of the competing influences of cell division, which
obviously increases the number of proliferative cells, and the general trend
of telomere shortening, we showed that there is a phase transition between a
normal state where a cell culture becomes senescent to a new state where a
cell culture can become immortal.
  
From our theory, we were able to answer the basic questions posed in the
introduction.  Specifically:
\begin{enumerate}
\item We determined the condition for whether a cell population ultimately
  becomes senescent or whether it continues to divide {\it ad infinitum}.
  The transition between these two regimes is given by the condition
  $\epsilon_M=1$, where $\epsilon_m=4Dk/Mv^2$ is a dimensionless measure of
  the relative effect of cell division, random fluctuations, and basal loss
  in the length evolution of a telomere.
\item We also found that for the case of $M=1$ telomere per cell, the mean
  time for a cell population to become senescent is
\begin{eqnarray*}
\langle t\rangle = \frac{x_0}{v}\frac{1}{\sqrt{1-\epsilon_1}}
\end{eqnarray*}
for $\epsilon_1<1$.  Here $x_0$ is the initial length of the telomere, $v$ is
the amount by which the telomere shrinks by basal loss in each division, and
$\epsilon_M \equiv 4Dk/Mv^2$ is a dimension measure of the relative
importance of cell division to basal loss.
\item Finally, we found that the total number of cells produced before the
  entire cell culture becomes senescent becomes extremely large as $\epsilon$
  approaches its critical value from below, as presented in Eq.~\eqref{Nsen}.
\end{enumerate}

Our results may be helpful for understanding how ALT cells maintain their
telomeres.  In these cells, increased T-SCE rate, and wide telomere size
distribution, and increased cell lifetimes have been observed \cite{TCE}.
These observations are all natural outcomes from our model.  How these
results depend on the number of short telomeres and on the details of the
T-SCE process are very important questions that are under investigation.
 
\acknowledgments{ Much of this work was performed when SR was on leave at the
  CNLS at Los Alamos National Laboratory.  He thanks the CNLS for its
  hospitality and P. Ferrari for encouragement.  KBB thanks E.H.~Goodwin for
  the useful discussions.  We also gratefully acknowledge financial support
  from NIH grant R01GM078986 (TA) and Jeffrey Epstein for support of the
  Program for Evolutionary Dynamics at Harvard University, DOE grant
  DE-AC52-06NA25396 (KB \& ST), and NSF grant DMR0535503 and DOE grant
  W-7405-ENG-36 (SR).}

\appendix*

\section{First-Passage for Biased Diffusion}

We recall some basic results about first-passage for biased diffusion on the
positive half line $x>0$ \cite{fpp} that will be used to describe cell
proliferation statistics.  According to Eq.~\eqref{def}, the length of each
telomere undergoes biased diffusion, in the continuum limit, with positive
bias ($v>0$) defined to be directed towards smaller telomere length $x$.  An
absorbing boundary condition at the origin imposes the constraint that when
$x$ reaches zero the cell effectively becomes senescent and is thus removed
from the population of dividing cells.

Let $c(x,t)$ be the concentration of diffusing particles at $x$ at time $t$.
The concentration evolves by the convection diffusion equation
\begin{equation}
\label{diff-app}
\frac{\partial c(x,t)}{\partial t} =  v\frac{\partial
  c(x,t)}{\partial x} + D \frac{\partial^2 c(x,t)}{\partial x^2}\,,
\end{equation}
For the initial condition $c(x,t=0)=\delta(x-x_0)$, corresponding to a single
particle starting at $x_0$, the concentration at any later time is \cite{fpp}
\begin{equation}
\label{cxt}
\begin{split}
c(x,t)=\frac{1}{\sqrt{4\pi Dt}}\,\Big[&e^{-(x-x_0+vt)^2/4Dt} \\
& -e^{+vx_0/D}\,e^{-(x+x_0+vt)^2/4Dt}\Big].
\end{split}
\end{equation}
The second term represents the ``image'' contribution; notice that the bias
velocity of the image is in the same direction as that of the initial
particle.  The exponential prefactor in the image term ensures that the
absorbing boundary condition $c(x=0,t)=0$ is always fulfilled.

From this concentration profile, the first-passage probability, namely, the
probability for a diffusing particle to hit the origin for the first time at
time $t$, is 
\begin{equation}
\label{F}
F(t)=D\frac{\partial c}{\partial x}-vc\,\,\Big|_{x=0} =
\frac{x_0}{\sqrt{4\pi Dt^3}}\,\, e^{-(x_0-vt)^2/4Dt}.
\end{equation}
The convective contribution to the flux, $-vc$, gives no contribution because
$c=0$ at $x=0$.  Until the particle hits the boundary it stays in the system;
hence its survival probability is simply
\begin{equation}
\label{S}
\begin{split}
S(t) \!=\! \int _0^\infty\!\! \!\!c(x,t)\, dx
= \frac{1}{2} \Bigg[&{\rm erfc}\Big(\frac{vt-x_0}{\sqrt{4Dt}}\Big)\\
&-e^{vx_0/D}{\rm erfc}\Big(\frac{vt+x_0}{\sqrt{4Dt}}\Big)\Bigg],
\end{split}
\end{equation}
where ${\rm erfc}(z)$ is the complementary error function. Since there is
only one absorbing point in the system, the survival probability and the
first passage probability are related by
\begin{equation}
\label{FS}
F(t)=- \frac{dS(t)}{dt}\,.
\end{equation}
From the asymptotics of the error function \cite{AS}, ${\rm erfc}(z)\sim
e^{-z^2}/\sqrt{\pi}\, z$, the long-time behavior of $S(t)$ is given by
\begin{eqnarray}
\label{Slarget}
S(t) &\sim& \sqrt{\frac{Dt}{\pi}} \frac{2x_0}{(vt)^2-x_0^2}\,\, e^{kt}\,\, 
e^{-(vt-x_0)^2/4Dt}\nonumber \\
&\sim & \sqrt{\frac{D}{\pi}} \frac{2x_0}{v^2}~ e^{vx_0/2D}~ t^{-3/2}~ e^{-(v^2/4D)t}.
\end{eqnarray}
As expected intuitively, the survival probability asymptotically decays
exponentially with time because the bias drives the particle towards the
absorbing point.  This result is used in Eqs.~\eqref{N} \& \eqref{Nt1} to
determine the number of active cells as a function of time.

\end{document}